\documentclass{article}
\renewcommand{\epsilon}{\varepsilon}
\begin{document}
\title{Conformal Symmetry and Cosmological Entropy Production}
\author{Winfried Zimdahl\footnote{Electronic address:
winfried.zimdahl@uni-konstanz.de}\\
Fachbereich Physik, Universit\"at Konstanz\\ PF M678, 
D-78457 Konstanz, Germany\\
and\\
Alexander B. Balakin\footnote{Electronic address: dulkyn@bancorp.ru}\\ 
Department of General Relativity and Gravitation  \\
Kazan State University, 420008 Kazan, Russia}

\date{\today}
\maketitle

\begin{abstract}
Introducing an effective refraction index of an isotropic cosmic 
medium, we investigate the cosmological fluid dynamics which is 
consistent with a conformal, timelike symmetry of a corresponding ``optical'' metric. 
We demonstrate that this kind of symmetry is compatible with the existence of a negative viscous pressure and, consequently, with 
cosmological entropy production. 
We establish an exactly solvable model according to which the viscous 
pressure is a consequence of a self-interacting one-particle force 
which is self-consistently exerted on the microscopic 
particles of a relativistic gas. 
Furthermore, we show that a sufficiently high decay rate of the 
refraction index of an ultrarelativistic cosmic medium results in an 
inflationary expansion of the universe. 
\end{abstract}  

\section{Introduction}

Standard relativistic cosmology relies on the cosmological principle 
according to which the Universe is spatially homogeneous and isotropic 
on sufficiently large scales. 
This symmetry requirement largely fixes the spacetime metric which may 
be written in the Robertson-Walker form (see, e.g., \cite{Wein})
\begin{equation}
\mbox{d}s ^{2} = - \mbox{d}t ^{2} + a ^{2}\left(t \right)
\left[\frac{\mbox{d}r ^{2}}{1-kr ^{2}} + r ^{2}\mbox{d}\theta ^{2} 
+ r ^{2}\sin ^{2}\theta \mbox{d}\phi ^{2}\right]\ .
\label{1}
\end{equation}
The quantity $k$ can take the values $k= 0, \pm 1$ and characterizes 
the three possible cases of the spatial curvature. 
Only one unknown function, 
the scale factor $a$, has to be determined by the gravitational field 
equations. 

For many purposes the cosmic substratum may be modeled as a fluid. 
So called fundamental observers are observers which are at rest with 
respect to the mean motion of the cosmic fluid. 
The symmetry requirements of the cosmological principle restrict the 
corresponding effective energy momentum tensor 
$T ^{ik}_{\left({\it eff} \right)}$ to be of the structure
\begin{equation}
T ^{ik}_{\left({\it eff} \right)} = \rho  u ^{i}u ^{k} + P h ^{ik}\ .
\label{2}
\end{equation}
Here, $\rho $ is the energy density measured by an observer comoving 
with the fluid four-velocity $u ^{i}$ which is normalized by $u ^{a}u _{a} = -1$. 
The quantity   
$h _{ik} \equiv  g _{ik} + u _{i}u _{k}$ is the spatial projection 
tensor with $h _{ik}u ^{k} = 0$. 
The total pressure $P$ is the sum  
\begin{equation}
P = p + \Pi 
\label{3}
\end{equation}
of an equilibrium part $p>0$ and a non-equilibrium part $\Pi \leq 0$  
which is connected with entropy production. 
A perfect fluid is characterized by $\Pi =0$. 
A scalar, viscous pressure is the only entropy producing phenomenon  
which is compatible with the symmetry requirements of the cosmological 
principle. 
Neither heat flows nor shear stresses are reconcilable with the 
assumption of spatial homogeneity and isotropy. 
This demonstrates that symmetries are generally connected with 
conditions on the properties and on the dynamics of the matter 
which generates this spacetime. 

Spacetime symmetries are invariantly characterized by symmetry groups. 
Isometries of a spacetime are described by Killing vectors (KVs). 
To characterize spacetimes that admit a conformal group one uses 
conformal Killing vectors (CKVs).  
A CKV $\xi ^{a}$ of the metric $g _{ik}$ is defined by 
\begin{equation}
\pounds _{\xi } g _{ik} \equiv  \xi _{i ;k} + \xi _{k ;i} 
= {\rm 2}\phi \left(x \right) g _{ik}\ ,
\label{4}
\end{equation}
where $\pounds _{\xi }$ is the Lie derivative along the vector field $\xi ^{i}$.   
The semicolon denotes the covariant derivative with respect to $g _{ik}$. 
For the special case $\phi =0$ the vector $\xi $ is a KV. 
Maximally symmetric spaces have the maximally possible number of KVs. 
The Minkowski space, e.g., has 10 KVs corresponding to its well-known 10 
symmetries and conservation laws.  
According to the cosmological principle, the four-dimensional spacetime contains 
maximally symmetric three-dimensional subspaces of constant curvature, 
corresponding to the spatial symmetries of homogeneity 
and isotropy, characterized by spacelike KVs. 
Timelike symmetries are not generally expected to exist in the expanding 
universe since a timelike KV characterizes a stationary spacetime. 
However, it is known \cite{TauWei} that under certain circumstances a 
{\it conformal}, timelike symmetry is possible in a Friedmann-Lema\^{\i}tre-Robertson-Walker 
(FLRW) universe. 
The existence of timelike or conformal, timelike KVs is closely related to 
``global'' equilibrium properties of perfect fluids. 
For a gaseous fluid system of particles with mass $m>0$, the 
global equilibrium condition can only be satisfied if the quantity 
$\xi ^{a}\equiv  u ^{a}/T$, where 
$T$ is the fluid temperature, is 
a (timelike) KV, equivalent to a stationary spacetime.  
For massless particles on the other hand, the condition for global 
equilibrium requires 
$u ^{a}/T$ to be a CKV. 
A CKV is compatible with the cosmological expansion.        
Macroscopically, a gas of massless particles obeys the equation of 
state for radiation, 
$p = \rho /3$. 
Only for this perfect fluid equation of state a global equilibrium 
is possible in the expanding universe. (The fact that there exists 
an equilibrium also in the non-relativistic limit is not relevant in this paper.)  
Any deviation from 
$p=\rho /3$ will destroy the conformal symmetry. 
In other words, the conformal symmetry singles out perfect fluids 
with the equation of state for radiation. 
Apparently, this also implies that a conformal symmetry is incompatible 
with a 
non-vanishing entropy production. 

The purpose of the present paper is to show that, contrary to this 
expectation, entropy production due to a scalar viscous pressure 
may well be consistent with a conformal, timelike  symmetry, 
albeit not a symmetry of the spacetime metric itself but the 
symmetry of an associated ``optical'' metric, characterized 
by a time dependent refraction index. 
The introduction of an effective refraction index of the 
medium as a new scalar parameter to characterize the cosmological 
evolution was shown to be useful in the context of cosmological 
particle production \cite{ZiBa01}. 
Here we explore the role of this parameter in more detail 
and for a different kind of non-equilibrium states for which the 
(not necessarily small) deviations from equilibrium are due to an increase 
in the entropy per particle,  
whereas the number of particles is preserved.  
In particular, we demonstrate that a time variation of 
the refraction 
index may take into account such kind of entropy producing deviations from the perfect 
fluid behavior.  
The standard global equilibrium for $p=\rho /3$ is then recovered as 
the limiting case for a refraction index unity. In this limit the 
optical metric coincides with the spacetime metric. 
Moreover, we show that a time dependent refraction index may be 
regarded as a consequence of the action of self-interacting forces on the  
microscopic constituents of an ultrarelativistic cosmic medium. 
Although entropy is produced, the microscopic particles are governed 
by an equilibrium distribution function. 
The deviations from standard equilibrium are mapped onto a non-zero, 
time varying chemical potential which is related to the refraction 
index in a simple way. Furthermore, we investigate the 
conditions under which the corresponding viscous pressure 
is sufficiently large to violate the strong energy condition and discuss 
simple examples for an inflationary evolution of the early universe.

The paper is organized as follows. 
In section 2 we summarize basic thermodynamic relations for a bulk viscous fluid. 
Section 3 presents the conformal symmetry concept of an optical metric 
and clarifies its relation to the production of entropy within the medium. 
Section 4 is devoted to a microscopic derivation of the fluid dynamics 
from relativistic gas theory in terms of a self-interacting one-particle force. 
Cosmological applications, in particular the conditions for power-law 
inflation, are discussed in section 5, while section 6 sums up our conclusions 
on the relation between cosmological entropy production and the 
conformal symmetry of an optical metric.   
Units have been chosen so that $c = k_{B} = h =1$.

\section{Basic thermodynamics}

We assume the cosmic substratum to be modeled by a homogeneous and 
isotropic fluid with the energy-momentum tensor (\ref{2}) 
and a particle number flow vector $N ^{i}$,  
given by 
\begin{equation}
N ^{i} = n u ^{i}\ ,
\label{5}
\end{equation}
where $n$ is the particle number density. 
The conservation laws for particle number and energy are 
\begin{equation}
N ^{i}_{;i} = 0 \quad\Rightarrow\quad\dot{n} + 3H n = 0 \ 
\label{6}
\end{equation}
and 
\begin{equation}
u _{i}T ^{ik}_{\left({\it eff} \right);k}\ = 0 
\quad\Rightarrow\quad \dot{\rho} + 3H\left(\rho + p + \Pi\right) = 0\ ,
\label{7}
\end{equation}
respectively, where $H \equiv  \dot{a}/a$ is the Hubble rate and 
$\dot{n} \equiv  n _{,i}u ^{i}$ etc. 
In the following we recall basic relations for the evolution of the 
relevant thermodynamical quantities. 
With the help of the Gibbs equation (see, e.g., \cite{Groot})
\begin{equation}
T \mbox{d}s = \mbox{d} \frac{\rho }{n} + p \mbox{d}\frac{1}{n}\ ,
\label{8}
\end{equation}
where $s$ is the entropy per particle and $T$ is the equilibrium 
temperature, we obtain 
\begin{equation}
n T \dot{s} = \dot{\rho } - \left(\rho + p \right)
\frac{\dot{n}}{n}\ .
\label{9}
\end{equation}
Using here the conservation laws (\ref{6})  and (\ref{7})  yields
\begin{equation}
n T \dot{s} = - 3H\Pi \ .
\label{10}
\end{equation}
Since $n$, $T$ and $H$ are positive, the entropy per particle increases 
for a negative viscous pressure $\Pi $.  
From the Gibbs-Duhem relation (see, e.g., \cite{Groot})
\begin{equation}
\mbox{d} p = \left(\rho + p \right)\frac{\mbox{d} T}{T} 
+ n T \mbox{d} \left(\frac{\mu }{T} \right)\ ,
\label{11}
\end{equation}
where $\mu $ is the chemical potential,  
it follows that 
\begin{equation}
\left(\frac{\mu }{T}\right)^{^{\displaystyle \cdot}} 
= \frac{\dot{p}}{nT} 
- \frac{\rho + p}{nT}\frac{\dot{T}}{T}\ .
\label{12}
\end{equation}
We assume  equations of state in the general form 
\begin{equation}
p = p \left(n,T \right)\ ,\ \ \ \ \ \ \ \ \ 
\rho = \rho \left(n,T \right)\ , 
\label{13}
\end{equation}
i.e., particle number density and temperature are taken as the independent 
thermodynamical variables. 
Differentiating the latter relation and using 
the balances (\ref{6}) and  
(\ref{7}) provides us with an evolution law for the temperature,  
\begin{equation}
\frac{\dot{T}}{T} = 
- 3H\Sigma \frac{\partial p}{\partial \rho }
\ ,
\label{14}
\end{equation}
where 
\begin{equation}
\Sigma \equiv  1  - \frac{\dot{s}}{3H } 
\frac{n}{\partial p/ \partial T}\ 
\label{15}
\end{equation}
and 
\[
\frac{\partial{p}}{\partial{\rho }} \equiv  
\frac{\left(\partial p/ \partial T \right)_{n}}
{\left(\partial \rho / \partial T \right)_{n}} \ ,
\ \ \ \ \ \ \ \ 
\frac{\partial{\rho }}{\partial{T}} \equiv  
\left(\frac{\partial \rho }{\partial T} \right)_{n}\ . 
\]
Moreover, we have used the general relation 
\[
\frac{\partial{\rho }}{\partial{n}} = \frac{\rho + p}{n} 
- \frac{T}{n}\frac{\partial{p}}{\partial{T}}\ ,
\] 
which follows from the fact that the entropy is a state function, i.e., 
$\partial ^{2}s/ \partial n \partial T 
= \partial ^{2}s/ \partial T \partial n$. 
For the special case of a perfect fluid, i.e., $\dot{s}=0$, and for 
ultrarelativistic matter, 
i.e., $p=\rho/3$, the temperature law (\ref{14}) specifies to 
\begin{equation}
\frac{\dot{T}}{T} = 
- \frac{\dot{a}}{a} \quad\Rightarrow\quad aT = {\rm const} \quad\quad 
\left(\dot{s}=0, \ p= \frac{\rho}{3}\right)\ ,
\label{16}
\end{equation}
which is the well-known behavior $T \propto a^{-1}$ for the radiation 
temperature in an 
expanding universe. 

With the help of Eqs. (\ref{13}), (\ref{6}), (\ref{9})  and (\ref{14}),  
the time evolution of the pressure may generally be written as     
\begin{equation}
\dot{p} = c _{s}^{2} \dot{\rho } 
- 3H\Pi  \left[\frac{\partial{p}}{\partial{\rho }} 
- c _{s}^{2}   \right]\ ,
\label{17}
\end{equation}
where 
\begin{equation}
c _{s}^{2}  = \left(\frac{\partial{p}}{\partial{\rho }} \right)_{ad} 
= \frac{n}{\rho + p}\frac{\partial{p}}{\partial{n}} 
+ \frac{T}{\rho + p} 
\frac{\left(\partial p / \partial T \right)^{2}}
{\partial \rho / \partial T}
\label{18}
\end{equation}
is the square of the adiabatic sound velocity $c _{s}$ [see, e.g. \cite{Weinberg}].  
Using Eq. (\ref{17})  in Eq. (\ref{12})  yields 
\begin{equation}
\left(\frac{\mu }{T} \right)^{^{\displaystyle \cdot}} 
= 3H \left\{\frac{\rho + p }{nT}
\left[\frac{\partial{p}}{\partial{\rho }} - c _{s}^{2} \right]
- \frac{\Pi }{nT}\left[\frac{\partial{p}}{\partial{\rho }}   
- \frac{\rho + p}{T \partial \rho / \partial T}\right]\right\} \ .
\label{19}
\end{equation}
\noindent
The relations so far are completely general. Below we shall use them in 
connection with a relativistic 
Maxwell-Boltzmann gas with internal self-interactions. 
It is expedient to emphasize that a viscous pressure $\Pi$ according to (\ref{19}) 
influences the time behavior of the quantity $\mu/T$. 
Lateron we shall be interested in a special case in which this non-equilibrium term is the only 
``source'' for $\left(\mu /T \right)^{\displaystyle \cdot}$.

\section{Conformal symmetry of optical metrics}

Given a spacetime metric $g _{ik}$ and an isotropic medium with a four-velocity $u ^{a}$ and a refraction index 
$n _{r}$, one introduces the ``optical'' metric $\bar{g}_{ik}$ by 
\cite{Gordon}
\begin{equation}
\bar{g}_{ik} \equiv   g _{ik} + \left(1- \frac{1}{n _{r}^{2}} \right)u _{i}u _{k} 
= h _{ik} - \frac{1}{n _{r}^{2}}u _{i}u _{k}\ . 
\label{20}
\end{equation}
\noindent
Optical metrics are known to be helpful in simplifying the equations of light 
propagation in isotropic, refractive media. 
With respect to a metric $\bar{g}_{ik}$ 
light propagates as in vacuum. 
Here we are interested in the role of optical metrics in relativistic gas dynamics. 
Guided by the connection between spacetime symmetries and equilibrium properties, outlined in the introduction, our starting point is to 
assume $\xi ^{a}\equiv  u ^{a}/T$  to be a conformal Killing vector 
of the optical metric, i.e., 
\begin{equation}
\pounds _{\xi }\bar{g}_{ik} = 2 \psi \bar{g}_{ik}\ .
\label{21}
\end{equation}
\noindent
The main objective of this paper is to explore the fluid dynamics which is 
compatible 
with the latter requirement. 
In particular, we are interested in the question to what extent the 
CKV condition (\ref{21}) admits the production of entropy in a cosmological 
context. 
In a first step, using the relations 
\[
\pounds _{\xi } \frac{u ^{i}}{T}={\rm 0} \ ,
\mbox{\ \ }\mbox{\ \ }
\pounds _{\xi } \frac{u _{i}}{T} 
= \frac{u ^{a}}{T}\pounds _{\xi } g _{ia}  
\ ,\quad 
\pounds _{\xi } S = \frac{\dot{S}}{T}\ ,
\]
the last one being valid for any scalar $S$, the left-hand side of 
Eq. (\ref{21}) is written as 
\begin{equation}
\pounds _{\xi }\bar{g}_{ik}  = \pounds _{\xi }g _{ik} 
+ 2 \frac{u _{i}u _{k}}{T}\left[\left(1- \frac{1}{n _{r}^{2}} \right)
\frac{\dot{T}}{T} 
+ \frac{\dot{n}_{r}}{n _{r}^{3}} \right] 
+ 2 \left[1- \frac{1}{n _{r}^{2}} \right]u _{k}u ^{a}
\pounds _{\xi }g _{ia}\ .
\label{22}
\end{equation} 
In a second step we find an explicit expression for the first term on the 
right-hand side 
of Eq. (\ref{22}). 
As any symmetric tensor, the Lie derivative of $g _{ik}$ may be decomposed 
into contributions 
parallel and perpendicular to the four-velocity. 
For homogeneous and isotropic media such a split amounts to 
\begin{equation}
\pounds _{\xi }g _{ik} \equiv \left(\frac{u _{i}}{T} \right)_{;k} 
+ \left(\frac{u _{k}}{T} \right)_{;i} 
= \frac{2}{T}\frac{\dot{a}}{a}g _{ik} 
+ \frac{2}{T}\left(\frac{\dot{T}}{T} + \frac{\dot{a}}{a}\right)
u _{i}u _{k}\ ,
\label{23}
\end{equation}
where we have used that in a FLRW universe 
$u ^{i}_{;k}=Hh^{i}_{k}\Rightarrow u ^{i}_{;i}=3 \dot{a}/a$. 
Combining Eqs. (\ref{22}) and (\ref{23})  
we obtain  
\begin{equation}
\pounds _{\xi }\bar{g}_{ik} = \frac{2}{T}\frac{\dot{a} }{a}h _{ik} 
+ 2 \frac{u _{i}u _{k}}{n _{r}^{2}T}
\left(\frac{\dot{n}_{r}}{n _{r}}  + \frac{\dot{T}}{T}\right) 
\label{24}
\end{equation}
for the left-hand side of Eq. (\ref{21}).    
Together with the definition (\ref{20}) one finds 
\[
\psi  = \frac{1}{T}\frac{\dot{a} }{a}\ 
\]
for the conformal factor in (\ref{21}) and 
\begin{equation}
\frac{\dot{n}_{r}}{n _{r}} = - \left[\frac{\dot{a}}{a} 
+ \frac{\dot{T}}{T}\right]\ 
\label{25}
\end{equation}
for the change rate of the refraction index. 
The last relation implies 
\[
n _{r} \propto \frac{1}{aT}\ , 
\]   
i.e., all  deviations from $aT = {\rm const}$ 
(pure radiation, $p=\rho /3$, cf. Eq. (\ref{16})) have been mapped onto 
the time-varying 
refraction index $n _{r}$.  
This demonstrates explicitly, that $n_r$ is indeed a useful quantity to 
characterize 
a fluid dynamics with a temperature evolution that is different from that for  
$\dot{s}=0$ and $p=\rho/3$. 

Inserting now the general temperature law (\ref{14})  into Eq. (\ref{25}), we obtain 
\begin{equation}
\frac{\dot{n}_{r}}{n _{r}} = - \left[1 - 3 \frac{\partial{p}}{\partial{\rho }} 
\Sigma\right]\frac{\dot{a}}{a}\ .
\label{26}
\end{equation}
The variable $\Sigma$ depends on $\dot{s}$ according to Eq. (\ref{15}). 
Consequently, Eq. (\ref{26}) 
establishes a relation between the rate of change of the refraction index 
and the expansion rate for a viscous cosmological fluid 
under the condition of a conformal symmetry of the 
optical metric $\bar{g}_{ik}$. 
This implies the statement that a conformal symmetry (\ref{21}) does not contradict a 
production of entropy. 
To obtain a better understanding of this feature, 
we present a derivation for such kind of fluid dynamics for 
a relativistic gas in the following section.  

\section{Kinetic theory}

The particles of a relativistic gas are assumed to move under the 
influence of a four-force $F ^{i}$ in between elastic binary collisions, described by 
Boltzmann's collision 
integral $C \left[f,f \right]$. 
The equations of motion of the gas particles are  
\begin{equation}
\frac{\mbox{d} x ^{i}}{\mbox{d} \gamma  } = p ^{i}\ ,
\ \ \ \ \ \ 
\frac{\mbox{D} p ^{i}}{\mbox{d} \gamma  } =  F ^{i}\ ,
\label{27}
\end{equation}
where $\gamma $ is a parameter along the particle worldline which for 
massive particles 
may be related to the proper time $\tau $ by $\gamma   = \tau /m $. 
Since the particle four-momenta are normalized according to 
$p ^{i}p _{i} = -m ^{2}$, the force $F ^{i}$ has to satisfy the relation 
$p _{i}F ^{i} = 0 $. 
The corresponding equation for the invariant one-particle distribution function 
$ f = f\left(x\left(\tau  \right),p \left(\tau  \right)\right)$ may be written as 
(cf. \cite{Groot,Ehl,IS,Stew}) 
\begin{equation}
L\left[f\right] +  \frac{\partial{\left(F ^{i}f \right)}}{\partial{p^{i}}}
= C [f,f]\ , 
\label{28}
\end{equation}
where 
\begin{equation}
L\left[f\right] \equiv
p ^{i}\frac{\partial{f}}{\partial{x ^{i}}} - \Gamma ^{k}_{il} p ^{i}p^{l}
\frac{\partial{f}}{\partial{p^{k}}} 
\label{29}
\end{equation}
is the Liouville operator. 
It will turn out that a suitably chosen effective one-particle force 
$F^i$ may give rise to a time-dependent refraction index on the macroscopic level. 
The particle number flow 4-vector 
$N^{i}$ and the energy momentum tensor $T^{ik}$ are
defined in a standard way (see, e.g., \cite{Ehl,IS}) as 
\begin{equation}
N^{i} = \int \mbox{d}Pp^{i}f\left(x,p\right) \mbox{ , } 
\ \ \ 
T^{ik} = \int \mbox{d}P p^{i}p^{k}f\left(x,p\right) \mbox{ .} 
\label{30}
\end{equation}
Since we shall identify the particle number flow in (\ref{30}) with the 
corresponding quantity in Eq. (\ref{5}), we have used here the same symbol, 
although these expressions may not coincide in the general case. 
The integrals in the definitions (\ref{30}) and in the following  
are integrals over the entire mass shell 
$p^{i}p_{i} = - m^{2}$. 
The entropy flow vector $S^{a}$ is given by \cite{Ehl,IS} 
\begin{equation}
S^{a} = - \int p^{a}\left[
f\ln f - f\right]\mbox{d}P \mbox{ , }
\label{31}
\end{equation}
where we have restricted ourselves to the case of 
classical Maxwell-Boltzmann particles. 
Using the general relationship \cite{Stew}
\begin{equation}
\left[\int p^{a_{1}}....p^{a_{n}}p^{b}f \mbox{d}P\right]_{;b} 
= \int p^{a_{1}}...p^{a_{n}}L\left[f\right] \mbox{d}P \ ,
\label{32}
\end{equation}
we find for the balance equations  
\begin{equation}
N ^{a}_{;a} = \int \mbox{d}P \left\{C \left[f,f \right] 
- \frac{\partial{\left(fF ^{i} \right)}}{\partial{p ^{i}}}\right\} = 0
\label{33}
\end{equation}
and
\begin{equation}
T^{ak}_{\ ;k} = \int \mbox{d}P p ^{a}\left\{C \left[f,f \right] 
- \frac{\partial{\left(fF ^{i} \right)}}{\partial{p ^{i}}}\right\} = 
\int \mbox{d}P f F ^{a}\ .
\label{34}
\end{equation}
In particular, the energy balance becomes 
\begin{equation}
u _{a}T^{ak}_{;k} = \int \mbox{d}P f u _{a}F ^{a}\ .
\label{35}
\end{equation}
The entropy production density is a sum of two terms: 
\begin{equation}
S ^{i}_{;i}  = \sigma _{c} + \sigma _{F}\ .
\label{36}
\end{equation}
Here, 
\begin{equation}
\sigma_{c} \equiv - \int \mbox{d}P 
C\left[f,f\right]  \ln f
\label{37}
\end{equation}
is the familiar contribution of Boltzmann's collision integral, while 
\begin{equation}
\sigma_{F} \equiv  - \int \mbox{d} P  
F ^{i} \frac{\partial{f}}{\partial{p ^{i}}}
= \int \mbox{d} P  
f \frac{\partial{F ^{i}}}{\partial{p ^{i}}}
\label{38}
\end{equation}
takes into account an entropy production due to the action of the force 
$F ^{i}$. 
Since Boltzmann's $H$ theorem guarantees $\sigma _{c} \geq 0$, 
we have 
\begin{equation}
S^i _{;i} - \sigma _{F} \geq 0 \ .
\label{39}
\end{equation}
\noindent
We restrict ourselves to the class of forces which admit  solutions 
of Eq. (\ref{28}) that are 
of the type of  J\"uttner's distribution function 
\begin{equation}
f^{0}\left(x, p\right) = 
\exp{\left[\alpha + \beta_{a}p^{a}\right] } \ ,
\label{40}
\end{equation}
where $\alpha = \alpha\left(x\right)$ and 
$\beta_{a}\left(x \right)$ is timelike. 
For $f \rightarrow f ^{0}$ the collision integral vanishes, i.e., 
$C\left[f ^{0},f ^{0}\right] = 0$.  
With $f$ replaced by $f^{0}$ in the definitions 
(\ref{30}) and (\ref{31}), the quantities $N^{a}$, $T^{ab}$ 
and $S^{a}$ 
may be split 
with respect to the unique four-velocity $u^{a}$ according to 
\begin{equation}
N^{a} = nu^{a} \mbox{ , \ \ }
T^{ab} = \rho u^{a}u^{b} + p h^{ab} \mbox{ , \ \ }
S^{a} = nsu^{a} \mbox{  .}
\label{41}
\end{equation}
Here we have identified the general fluid quantities of the previous section 
with those following from the dynamics of a Maxwell-Boltzmann gas. 
Note that the energy-momentum tensor $T ^{ik}$ in Eqs. (\ref{30}) 
and (\ref{41})  does {\it not} coincide with the tensor 
$T ^{ik}_{\left({\it eff} \right)}$ introduced in Eq. (\ref{2}).  
The exact integral expressions for $n$, $\rho$, $p$ are given, 
e.g., in \cite{Groot}).  
The entropy per particle $s$ is  
\begin{equation}
s = \frac{\rho + p}{nT} - \frac{\mu }{T} \ .
\label{42}
\end{equation}
For $f \to f ^{0}$ the energy balance (\ref{35}) takes the form 
\begin{equation}
\dot{\rho } + 3H \left(\rho + p \right) 
= - \int \mbox{d}P f ^{0}u _{a}F ^{a}\ .
\label{43}
\end{equation}
The entropy production density becomes 
\begin{equation}
S ^{a}_{;a} = \sigma _{F} = n \dot{s} = - \int \mbox{d}P\beta _{i}F ^{i}\ .
\label{44}
\end{equation}
With the identification $\beta _{i} = u _{i}/T$ we may write 
\begin{equation}
n \dot{s} = - 3H\frac{\Pi }{T}\ ,
\label{45}
\end{equation}
where we have introduced the quantity $\Pi $ by 
\begin{equation}
3H \Pi \equiv \int \mbox{d}Pu _{i}F ^{i}\ .
\label{46}
\end{equation}
Notice that we have identified this kinetic theory based viscous pressure 
with the corresponding phenomenological quantity of the previous sections. 
Consequently, the energy balance (\ref{43}) takes the form (\ref{7}).  
This implies the conclusion that 
{\it a non-vanishing viscous pressure may generally be compatible with an 
equilibrium-type distribution function, provided, the gas particles move in a 
suitable force field.}
Conventionally, a bulk viscous pressure arises as a consequence of deviations 
from an equilibrium distribution. Wheras these deviations have to be small in 
standard non-equilibrium theories, a corresponding restriction is absent 
in the present case. 

Substituting the distribution (\ref{40})  into Eq. (\ref{28}) we obtain
\begin{equation}
p^{a}\alpha_{,a} +
\beta_{\left(a;b\right)}p^{a}p^{b}   
=  - \beta _{i}F ^{i} - \frac{\partial{F ^{i}}}{\partial{p ^{i}}}
\mbox{ .} 
\label{47}
\end{equation} 
To make further progress, we have to chose specific force types. 
The most general force compatible with the cosmological principle is 
\cite{ZSBP}
\begin{equation}
F ^{i} = - F \left(Ep ^{i} + p^a p_a u ^{i} \right)\ ,
\label{48}
\end{equation}
where $E \equiv  -u _{i}p ^{i}$ is the particle energy measured by a 
comoving observer. Generally, $F$ may still depend on $E$. 
With 
\begin{eqnarray}
\frac{\partial{F ^{i}}}{\partial{p^{i}}} &=& 
- \frac{\partial{F }}{\partial{p ^{i}}} 
\left( Ep ^{i} + p^a p_a u^{i}\right) 
- F\left(-u_i p ^{i} + 4E + 2 p_i u ^{i}\right) \nonumber\\
&=& \frac{\partial{F}}{\partial{E}}
\left(m^{2} - E^{2}\right) -3 FE \ ,
\label{49}
\end{eqnarray}
the condition (\ref{47})  then specifies to 
\begin{equation}
p^{a}\alpha_{,a} +
\beta_{\left(a;b\right)}p^{a}p^{b}   
=  - 3F u _{a}p ^{a} 
+ \left(\frac{\partial{F}}{\partial{E}}
- \frac{F}{T}\right) 
h _{ab}
p^{a}p^{b} 
\mbox{ .} 
\label{50}
\end{equation} 
Now we restrict ourselves to the case $\partial F/ \partial E = 0$ and 
consider massless particles $p^i p _i = - m^2 = 0$, which 
macroscopically corresponds to an equation of state 
$p = \rho /3$.  
As a consequence we find 
\begin{equation}
\alpha _{,a} = - 3 F u _{a} 
\quad\Rightarrow\quad \dot{\alpha} = 3 F \ 
\label{51}
\end{equation}
and 
\begin{equation}
\beta _{\left(a;b \right)} \equiv  \frac{1}{2}
\left[\left(\frac{u _{a}}{T}\right)_{;b} 
+ \left(\frac{u _{b}}{T}\right)_{;a}
 \right]= \frac{H}{T}g _{ab} 
- \frac{F}{T}u _{a}u _{b}\ .
\label{52}
\end{equation} 
Combining the relations (\ref{51}) and (\ref{52}), we obtain 
\begin{equation}
\beta _{\left(a;b \right)} = 
\frac{1}{2}\pounds _{\xi }g _{ab} = 
\frac{H }{T}g _{ab} - \frac{\dot{\alpha}}{3T}u _{a}u _{b}\ .
\label{53}
\end{equation}
Using  the equation of state $p=nT=\rho /3$ in Eq. (\ref{12}) and identifying 
$\alpha=\mu /T$, we have 
\begin{equation}
\left(\frac{\mu }{T} \right)^{\displaystyle \cdot} = \dot{\alpha} 
= - 3 \left(\frac{\dot{a}}{a} + \frac{\dot{T}}{T} \right)\ .
\label{54}
\end{equation}
In the standard case (without a force contribution) one has 
$\alpha = {\rm const}\left(=0 \right)$ and recovers the familiar cooling rate 
$\dot{T}/T = - \dot{a}/a$ for a relativistic fluid. 
We conclude that 
{\it the force field manifests itself in the existence of a non-vanishing, 
time dependent chemical potential.} From (\ref{54}) and (\ref{53})  we recover 
\begin{equation}
Tu ^{a}u ^{b}\beta _{\left(a;b \right)} = 
\frac{T}{2}u ^{a}u ^{b}\pounds _{\xi }g _{ab} = 
\frac{\dot{T}}{T}\ ,
\label{55}
\end{equation}
which shows the consistency of our approach. 
Comparing (\ref{53})  with (\ref{23})  we obtain 
\begin{equation}
\frac{\dot{T}}{T} + \frac{\dot{a}}{a} = - \frac{\dot{\alpha }}{3}
\quad\Rightarrow\quad \frac{\dot{n}_{r}}{n _{r}} = \frac{\dot{\alpha }}{3} 
= F\ , 
\label{56}
\end{equation}
i.e., with the identification $\dot{n}_{r}/n _{r}= \dot{\alpha }/3$ 
we recover the conformal symmetry relation (\ref{25}). 
Moreover, we have traced back the time variation of the refraction index to 
a self-consistent interaction within the system. 
From the moment equation (\ref{34})  it follows that 
\begin{equation}
T ^{ak}_{\ ;k} = F u _{k}T ^{ak} 
\quad\Rightarrow\quad 
u _{a}T ^{ak}_{;k} = - \dot{\rho } - 3H\left(\rho +p \right) 
= F \rho  \ .
\label{57}
\end{equation}
Since $\rho =3nT$ we reproduce Eq. (\ref{7}) 
with 
\begin{equation}
3H \Pi = 3nTF 
\quad\Rightarrow\quad 
F = \frac{\Pi }{nT}H \ .
\label{58}
\end{equation}
This is consistent with (\ref{46}) and (\ref{48}).  
The quantity $F$ depends on macroscopic fluid quantities. 
According to (\ref{27}) and (\ref{48}), these macroscopic quantities 
determine the motion of the individual microscopic particles 
which themselves are the constituents of the macroscopic medium. 
Consequently, the force $F ^{i}$ describes a self-interaction of the fluid. 
This self-interaction is accompanied by a time variation of the chemical 
potential, equivalent to a time dependent refraction index. 

From the entropy production density (\ref{44}) we find 
\begin{equation}
S ^{a}_{;a} = - \frac{F}{T}u _{a}u _{b}T ^{ab} 
= - \frac{3H\Pi }{T} = n \dot{s}\ .
\label{59}
\end{equation}
This means 
\begin{equation}
\dot{s} = - \frac{3H \Pi }{nT}\ ,
\label{60}
\end{equation}
consistent with (\ref{45}) and the  phenomenological relation  (\ref{10}).  
The ratio $\dot{s}/\left(3H \right) = - \Pi / \left(nT \right)$ which 
characterizes the quantity $\Sigma$ is just the ratio between non-equilibrium 
pressure $\Pi $  and equilibrium pressure $p=nT$.  
Via Eqs. (\ref{51}) and (\ref{58}), relation (\ref{60})  also implies 
 
\begin{equation}
\left(\frac{\mu }{T} \right)^{\displaystyle \cdot} 
= \frac{3H \Pi }{nT} = 
-\dot{s} = 3 \frac{\dot{n}_{r}}{n _{r}} = 3F\ .
\label{61}
\end{equation}
With the first equation in (\ref{61}) we have recovered the special case of 
Eq. (\ref{19}) for 
$p=nT$, $\rho = 3nT$, $c _s^{2} = \partial p / \partial \rho =1/3$. 
Furthermore, we have precised, how a decaying refraction index generates a 
negative contribution to the pressure, equivalent to positive entropy production.  
Integrating the part 
$\left(\mu /T\right)^{\displaystyle \cdot} 
= 3 \dot{n}_{r}/n _{r}$
of Eq. (\ref{61}) 
by assuming $\mu /T \rightarrow 0$ for $n _{r}\rightarrow 1$ 
(standard radiation with $\mu =0$ and $n _{r}=1$), we find 
\begin{equation}
\exp{\left[\frac{\mu }{T} \right]} = n _{r}^{3}\ .
\label{62}
\end{equation}
\noindent
Consequently, the distribution function (\ref{40}) can be written as 
\begin{equation}
f ^{0} = \exp{\left[\frac{\mu }{T} \right]}
\exp{\left[-\frac{E}{T} \right]} 
= n _{r}^{3}\exp{\left[-\frac{E}{T} \right]}\ , 
\label{63}
\end{equation}
i.e., the dissipative force gives rise to a time varying factor $n _{r}^{3}$ 
in the equilibrium-type distribution function. 

In order to check the consistency of the entire framework, we may now 
use the force (\ref{48}) with 
(\ref{58}) in the microscopic equations of motion in Eq. (\ref{27}).  
Contracting the latter with the macroscopic four-velocity results in 
\begin{equation}
\frac{\mbox{D}\left(u _{i}p ^{i} \right)}{\mbox{d}\gamma  } 
= u _{i}F ^{i} + u _{i;k}p ^{i}p ^{k} \ ,
\label{64}
\end{equation}
where we have used that 
\[
\frac{\mbox{D} u ^{i}}{\mbox{d} \gamma } = u ^{i}_{;n}p ^{n}\ .
\]
Since $u_{i;k}=Hh_{ik}$ for the homogeneous, isotropic case under 
consideration here 
and with Eq. (\ref{48}) for $m=0$ and (\ref{58}) this reduces to 
\begin{equation}
-\frac{\mbox{d}E}{\mbox{d}\gamma  } 
= \left(1 + \frac{\Pi }{nT} \right)HE ^{2}\ .
\label{65}
\end{equation}
Since $\mbox{d}E/ \mbox{d}\gamma = E _{,m}p ^{m} = E\dot{E}$ 
[cf. Eq. (\ref{27})] we obtain finally 
\begin{equation}
-\frac{\dot{E}}{E} 
= \left(1 + \frac{\Pi }{nT} \right)\frac{\dot{a} }{a}
\quad\Rightarrow\quad 
\frac{\dot{E}}{E} = \frac{\dot{T}}{T}\ .
\label{66}
\end{equation}
It follows that the distribution function (\ref{63}) 
is indeed maintained for particles under the action of the force $F ^{i}$, 
which proves the consistency of our approach.   
While the second factor in the expression (\ref{63}) for $f ^{0}$ remains 
constant, the first one changes according to Eq. (\ref{61}).

\section{Conformal symmetric universe}

Now we combine the non-equilibrium fluid dynamics of the previous section 
with the gravitational field equations.
In a homogeneous and isotropic, spatially flat universe 
Einstein's equations reduce to   
\begin{equation}
8\pi G \rho = 3 H ^{2}\ ,
\mbox{\ \ \ \ }
\dot{H} = - 4\pi G\left(\rho + p + \Pi  \right)\ .
\label{67}
\end{equation} 
\noindent
The relation between $\Pi$ and $\dot{n}_{r}$ in Eq. (\ref{61}) may be solved 
with respect to the non-equilibrium 
pressure,  
\begin{equation}
\Pi = \frac{1}{3}\frac{\dot{n}_{r}/n _{r}}{\dot{a}/a}\rho \ .
\label{68}
\end{equation} 
For $\rho +3P$ we obtain 
\begin{equation}
\rho + 3P =  \frac{1}{3} \left(\frac{\dot{n}_r /n_r}{\dot{a}/a} + 2 \right)
\ .
\label{69}
\end{equation}
The case $\rho +3P=0$ is equivalent to 
\begin{equation}
- \frac{\dot{n}_{r}}{n _{r}} 
= 2 \frac{\dot{a}}{a}\ .
\label{70}
\end{equation}
Consequently, for a decaying refraction index with 
$|\dot{n}_{r}/n _{r}| \geq 2\dot{a}/a$ 
the strong energy condition $\rho +3P > 0$ is violated 
and we have accelerated expansion. 
This may be explicitly demonstrated for a simple example. 
With $\Pi$ from (\ref{68})  
the energy balance (\ref{7}) may be written 
\begin{equation}
\dot{\rho } + \left(4 \frac{\dot{a}}{a} 
+ \frac{\dot{n}_{r}}{n _{r}} \right)\rho  =0 
\quad\Rightarrow\quad \rho n _{r}a ^{4} = {\rm const}\ .
\label{71}
\end{equation}
The simplest case is to assume a decay rate which is proportional to the Hubble rate: 
\begin{equation}
\frac{\dot{n}_{r}}{n _{r}} = - 4\beta \frac{\dot{a}}{a} 
\quad\Rightarrow\quad \rho a ^{4 \left(1-\beta  \right)} = {\rm const}\ .
\label{72}
\end{equation} 
The constant factor $\beta$ in the range $0 \leq \beta < 1$ has been chosen 
for convenience. 
Of course, $\beta $ can be assumed constant at most piecewise. 
Integration of Friedmann's equation yields 
\begin{equation}
a \propto t ^{\frac{1}{2 \left(1-\beta  \right)}}\ .
\label{73}
\end{equation}
Obviously, there is accelerated expansion $\ddot{a}\geq 0$ for  
$\beta \geq 1/2$, which is consistent with (\ref{70}). 
The limit $\beta = 0$ corresponds to the standard radiation 
dominated universe. 
If $\beta $ approaches $\beta \approx 1$, 
we have $\rho \approx {\rm const}$ and, according to Eq. (\ref{67}),   
an approximately constant Hubble rate $H$, 
i.e., exponential expansion. 
Any $\beta $ in the range $0 \leq \beta < 1$ is connected with entropy production. 
Combining (\ref{68}) and (\ref{72})  we find 
\begin{equation}
\Pi = - \frac{4}{3}\beta \rho 
\label{74}
\end{equation}
and, with (\ref{60}),   
\begin{equation}
\dot{s} = 12 \beta \frac{\dot{a}}{a} 
= 6 \frac{\beta }{\left(1-\beta  \right)t} 
\ .
\label{75}
\end{equation}
The entropy per particle grows logarithmically with the cosmic time. 
According to (\ref{58}), (\ref{74}), and (\ref{48}), the corresponding force is 
\begin{equation}
F ^{i} = 4 \beta HE p ^{i}\ .
\label{76}
\end{equation}
The action of this force on the microscopic constituents of the medium 
realizes a power-law evolution (\ref{73}) of the universe, which is a 
manifestation of a conformal symmetry (\ref{21}) of the optical metric 
(\ref{20}) and, moreover, implies a positive entropy production according 
to (\ref{75}).  
This represents an exactly solvable model, both macroscopically and 
microscopically, of a specific non-equilibrium configuration of a 
self-interacting cosmic gas.

\section{Conclusions}

We have shown that the introduction of a refraction index $n_r$ 
of the cosmic medium 
as a new scalar parameter allows us to understand early phases of  
the cosmological evolution as a specific 
non-equilibrium 
configuration of a self-interacting relativistic gas. 
The latter is characterized by an equilibrium-type distribution 
function of the microscopic particles. Deviations from equilibrium 
are mapped onto a time-varying chemical potential which is related to 
$n _{r}$ in a simple way. 
The refraction index defines an optical metric 
$\bar{g}_{ik}=g_{ik}-\left(1-n_r ^{-2}\right)u_i u_k$ 
which coincides with the spacetime metric  
$g_{ik}$ for $n_r =1$. 
A decaying refraction index gives rise to a negative contribution 
to the scalar pressure of the cosmic substratum. 
We have established an exactly solvable, self-consistent model 
for this type of dynamics which is connected with entropy production, 
but at the same time realizes a conformal, timelike symmetry of 
the optical metric.  
In other words, for any deviation from the behavior of an ultrarelativistic  perfect fluid which is due to a scalar, 
viscous pressure,  
one may find a refraction index such that the  
evolution of the universe appears as the manifestation of a conformal, 
timelike symmetry of the corresponding optical metric. 
For a decay rate of the refraction index which is of the order of the 
Hubble expansion rate, the associated  
negative pressures may lead to a violation of the strong energy 
condition $\rho + 3P >0$.  
Consequently, the resulting dynamics is inflationary. \\
\ \\
{\bf Acknowledgment}\\
This paper was supported by the Deutsche Forschungsgemeinschaft and by NATO.\\

\end{document}